# Ferromagnetic Double Perovskite Semiconductors with Tunable Properties

*Lun Jin\*, Danrui Ni, Xin Gui, Daniel B. Straus, Qiang Zhang and Robert J. Cava\**

L. Jin, D. Ni, X. Gui, D. B. Straus and R. J. Cava

Department of Chemistry, Princeton University, Princeton, NJ, 08544, USA

**\*** E-mails of corresponding authors: ljin@princeton.edu; rcava@princeton.edu

Q. Zhang

Neutron Scattering Division, Oak Ridge National Laboratory, Oak Ridge, TN, 37831, USA

We successfully dope the magnetically silent double perovskite semiconductor $Sr_2GaSbO_6$ to induce ferromagnetism and tune its band gap, with $Ga^{3+}$ partially substituted by the magnetic trivalent cation $Mn^{3+}$, in a rigid cation ordering with $Sb^{5+}$. Our new ferromagnetic semiconducting $Sr_2Ga_{1-x}Mn_xSbO_6$ double perovskite, which crystallizes in tetragonal symmetry (space group *I*4/*m*) and has tunable ferromagnetic ordering temperature and band gap, suggests that magnetic ion doping of double perovskites is a productive avenue towards obtaining materials for application in next-generation oxide-based spintronic devices.

## 1. Introduction

Magnetic semiconductors combine ferromagnetic and semiconducting properties and are therefore promising candidates for applications in next generation spintronic devices.[1,2] If such materials are applied, then the processing, communication and mass storage of information can be achieved simultaneously.[3,4] Given that oxide perovskites are one of the largest and most useful families of materials, and are of specific interest for use in next generation devices,[5–7] the discovery of perovskite-family ferromagnetic oxide semiconductors is a high-potential field of research. There are only a few existing experimental or theoretical examples of such materials[8–11], and the importance of exploring tunable ferromagnetic semiconductors in the oxide perovskite family cannot be over-emphasized.

Ferromagnetism and semiconducting properties have been known to coexist in Eu- and Mn-containing chalcogenides[12–17], in spinels[18–20] and in main-group/manganese-blended II-VI and III-V compounds[21–26] for decades, leading to the observation of favorable properties associated with the interactions between these two phenomena. In addition to a variety of other materials that have been explored for this application in recent years[27], low-dimensional chalcogenide and halide compounds (e.g. $Cr_2Ge_2Te_6$, $CrI_3$ and $VI_3$) have recently been of interest,[27–32] although integrating these materials with oxide-based electronics will be challenging. Despite the fact that a considerable number of magnetic semiconductors have been discovered until now, perovskite-based spintronic materials remain rare. Oxide perovskites are of special interest as they are particularly integrable into next-generation microelectronic devices, and recent reports about developing halide-perovskite-based spintronics[33,34] indicate that it may be appropriate to revisit the search for magnetic semiconductor materials in the oxide perovskite family as well. Oxide perovskites have been studied intensively for decades and represent one of the major classes of applied materials, yet ferromagnetic semiconducting oxide perovskites are rather unusual. This could be explained by the Goodenough-Kanamori rules[35], which indicate the coupling between equivalently filled metal ion orbitals in perovskite-like structures will be antiferromagnetic, thus the vast majority of semiconducting oxide perovskites containing only one type of transition metal ion are antiferromagnets.

The preparation of cation-ordered double perovskites can be daunting, because such materials are often thermodynamically unfavorable with respect to simple disordered solid solutions, unless the cation ordering is driven by strong chemical factors.[8] We postulate that a semiconducting $A_2BB'O_6$ B-cation-ordered double perovskite that lacks magnetic transition metal cations in the lattice, but displays a suitable semiconducting band gap, can serve as an excellent host material for making a dilute ferromagnetic semiconductor. The ferromagnetism would then be introduced into the semiconductor by partially substituting magnetic transition metal centers for one of the non-magnetic B-cations. $Sr_2GaSbO_6$ is one such semiconducting B-site cation ordered double perovskite.[36] Since $Sr_2MnSbO_6$[37,38] is a well-established compound, the magnetic transition element ion $Mn^{3+}$ was chosen to partially substitute for $Ga^{3+}$ in $Sr_2GaSbO_6$ to attempt to induce ferromagnetism. Thus in this study we have prepared a series of doped $A_2BB'O_6$ double perovskite phases $Sr_2Ga_{1-x}Mn_xSbO_6$ ($0.1 \leq x \leq 0.9$) and have fully characterized them from the structural, magnetic and optical points of view. We have found that the $Mn^{3+}$ dopant yields both tunable band gaps and ferromagnetism. The findings reported here are significant since ferromagnetic semiconducting double perovskites with tunable properties are extremely rare and to the best of the authors' knowledge the current series is unique.

## 2. Results and Discussion

### 2.1 Structural Characterization

Sr$_2$GaSbO$_6$ is a B-site cation ordered double perovskite with the usual double perovskite arrangement of alternating BO$_6$-B'O$_6$ octahedra, crystallizing in tetragonal symmetry. Laboratory X-ray powder diffraction (XRD) data collected from Sr$_2$GaSbO$_6$ powder is well fit by a body centered tetragonal unit cell, with lattice parameters $a$ = 5.54 Å and $c$ = 7.90 Å, in good agreement with the literature[39]. To further investigate the crystal structure, especially the oxygen positions, which cannot be determined well from XRD, we performed powder neutron diffraction measurements. The refinement on the time of flight (TOF) neutron powder diffraction data against the reported structural model converged smoothly and provided a satisfactory agreement parameter ($wR$ = 5.100%; $GOF$ = 3.83). Observed, calculated and difference plots from the Rietveld refinement of the B-site -ordered structure of the double perovskite Sr$_2$GaSbO$_6$ (space group $I4/m$) against the neutron powder diffraction data are shown in the Supporting Information (SI) Figure S1. The detailed structural parameters and crystallographic positions are presented in Table S1.

The initial attempt was carried out by 10% Mn-doping. Detailed structural characterization of Sr$_2$Ga$_{0.9}$Mn$_{0.1}$SbO$_6$ was performed by refining ambient temperature TOF neutron powder diffraction data against the structure of the parent phase Sr$_2$GaSbO$_6$[39], with the all-Ga$^{3+}$ site replaced by 90% Ga$^{3+}$ and 10% Mn$^{3+}$. The refinement converged smoothly and provided a satisfactory agreement parameter ($wR$ = 4.179%, $GOF$ = 3.54). Observed, calculated and difference plots from the Rietveld refinement of Sr$_2$Ga$_{0.9}$Mn$_{0.1}$SbO$_6$ (space group $I4/m$) against the ambient neutron powder diffraction data are shown in **Figure 1**a, and the structure is depicted in Figure 1b. The detailed structural parameters and crystallographic positions are presented in **Table 1**. Selected bond lengths for the (Ga/Mn)O$_6$ octahedron, the SbO$_6$ octahedron and the SrO$_{12}$ polyhedron are listed in Table S2.

The successful synthesis of 10% doped single phase $Sr_2Ga_{0.9}Mn_{0.1}SbO_6$ as an ordered $A_2BB'O_6$ double perovskite indicates that the undoped parent phase $Sr_2GaSbO_6$ can form at least a partial double perovskite solid-solution with $Sr_2MnSbO_6$, so an investigation of the remainder of the phase diagram was carried out. We found that $Sr_2Ga_{1-x}Mn_xSbO_6$ exhibits a full-range tetragonal ordered B site solid solution ($0 \leq x \leq 1$), most simply illustrated by continuous changes in the lattice parameters upon substituting $Ga^{3+}$ with $Mn^{3+}$. Laboratory powder XRD patterns for the $Sr_2Ga_{1-x}Mn_xSbO_6$ ($x$ = 0.1, 0.3, 0.5, 0.7 and 0.9) samples are stacked in **Figure 2**a to show the continuous shift of the diagnostic tetragonal (200 / 020) doublet around a 2θ position of 46°. The absence of the pseudo-cubic peak indicates that the BB' cation ordering is conserved through the whole doping range.

Lattice parameters were determined by Rietveld refinements of the laboratory XRD data. The resulting $a$, $c$ and cell volume $V$ values are plotted against the doping level $x$ in $Sr_2Ga_{1-x}Mn_xSbO_6$ in Figure 2b & c. The data for the end-member $Sr_2MnSbO_6$[40] are extracted from the literature. The lattice parameter $a$ exhibits an upward-opening parabolic curvature (minimum at $x$ = 0.4) while $c$ exhibits a downward-opening one (maximum at $x$ = 0.8). The cell volume $V$ possesses a linear relationship with increasing $x$. The reported data for the $Sr_2MnSbO_6$ end-member[40] does not fall in line with the materials reported here. One potential explanation for this observation is that the all-Mn end member $Sr_2MnSbO_6$ may be in proximity to a structural transition, the determination of which is beyond the scope of the current study.

Finally, scanning electron microscope (SEM) images were collected to observe the surface morphologies of powder particles of the Mn-doped double perovskites. Representative images are shown in Figure S2. The grain size tends to decrease at the elevating doping levels. The perovskite phase is clearly maintained in single phase form for these materials – no impurity particles were detected. The equipped energy-dispersive X-ray spectroscope (EDX) provided compositional results that are consistent with the nominal ones.

## 2.2 Physical Property Characterization

*2.2.1. Magnetic Properties*

The temperature-dependent and field-dependent magnetization data collected for the $Sr_2Ga_{1-x}Mn_xSbO_6$ series and the Mn-doped single perovskite $LaGa_{1-x}Mn_xO_3$ series, prepared for comparison, are plotted as magnetic susceptibility $\chi$ ($M/H$) against temperature $T$ and magnetic moment $M$ against field $H$, in the Supporting Information (SI) Figures S3 – S16. The magnetic susceptibility data, over the suitable temperature range (selected as the straight-line part of the 1/$\chi$ vs. T curves) of the doped materials were fitted to the Curie-Weiss law ($\chi = C/(T - \theta) + \chi_0$), to yield the Curie constants $C$ and Weiss temperatures $\theta$ that are listed in Table S3 and depicted in **Figure 3**a & b. The magnetization data reported for the end-member $Sr_2MnSbO_6$[38] fall in line with the materials reported in this study.

For $x = 0.1$ in the $Sr_2Ga_{1-x}Mn_xSbO_6$ series, the inverse of $\chi$ exhibits a perfect linear relationship with temperature $T$ in nearly the whole measured temperature range $1.8 \leq T / K \leq 300$, and the fitted Curie constant coincides with a high-spin $d^4$ electronic configuration for $Mn^{3+}$. The Weiss temperature $\theta$ has a positive value of + 3.43(9) K, which suggests that a ferromagnetic interaction weakly dominates in this material. With increased $Mn^{3+}$ doping concentration, a local maximum in the magnetic susceptibility curve starts being observed ($x = 0.3$) within our measurable temperature regime ($T \geq 1.8$ K). For $x = 0.4$, the Curie constant starts to deviate from what is expected for the spin-only value and with further Mn-doping ($x \geq 0.4$), clear divergence starts to appear between the zero-field-cooled (ZFC) and field-cooled (FC) curves in plots of magnetic susceptibility $\chi$ ($M/H$) against temperature $T$. The Weiss temperatures $\theta$ are all positive and increase in magnitude upon elevating the $Mn^{3+}$ doping level (Figure 3b). Furthermore, the value of the ferromagnetic ordering temperature $T_c$ (estimated as the temperature at which the $1/\chi$ vs. T curve deviates from Curie-Weiss law linearity. Due to the character of the $\chi$ vs. T data, an exact measure of Tc is not possible by the methods employed.) for the $Sr_2Ga_{1-x}Mn_xSbO_6$ series increases systematically with increasing $x$. The isothermal magnetization data collected from the Mn-doped materials at 300 K as a function of applied field is linear and passes through the origin, with a small positive slope, while analogous data collected at 2 K have positive slopes as well, but start to exhibit easily observed hysteresis at 30% Mn-doping, serving as strong evidence of ferromagnetic behavior, consistent with the temperature-dependent magnetization data.

Although the Mn-doped disordered single perovskite $LaGa_{1-x}Mn_xO_3$ series has been well-studied[41–45], samples of this series in the composition range $0.1 \leq x \leq 0.5$ were characterized magnetically as a comparison to the Mn-doped double perovskite. The magnetization data were scaled to their equivalent double perovskite values (i.e. for $La_2Ga_{2-2x}Mn_{2x}O_6$) for consistency. The magnetic susceptibility data can be fitted to the Curie-Weiss law in suitable temperature regimes ($100 \leq T / K \leq 300$ to $200 \leq T / K \leq 300$ depending on the value of $x$). The resulting Curie constants follow a similar trajectory as their double perovskite analogs but in a more linear fashion (Figure 3a), while the magnitude of the Weiss temperature is almost doubled in the single disordered perovskite series at identical Mn concentrations (Figure 3b). The isothermal magnetization data at 300 K and 2 K are consistent with the temperature-dependent magnetization data and are in good agreement with the literature.[41,43,45] The magnetization data collected from double- and single-perovskites with 40% total Mn-concentration on the B-sites ($LaGa_{0.6}Mn_{0.4}O_3$ and $Sr_2Ga_{0.2}Mn_{0.8}SbO_6$) are stacked in Figure 3c & d, as representative data. The temperature-dependent magnetic susceptibility curves coincide with each other in the temperature range $150 \leq T / K \leq 300$, while they differ significantly at lower temperatures.

In **Figure 4**, (a) the observed effective moment per formula unit $\mu_{eff.obs}$ is plotted against the calculated effective moment per formula unit $\mu_{eff.cal}$ based on the spin-only value of the transition metal magnetic moment, (b) and the observed magnetization $M$ per formula unit collected at $T = 2$ K under an applied field of 9 T is plotted against the calculated saturation magnetization $M_s$ per formula unit. The suitably scaled disordered single perovskite $LaGa_{1-x}Mn_xO_3$ series is included for comparison. For both Mn-doped series, the observed $\mu_{eff.obs}$ only agrees with the spin-only $\mu_{eff.cal}$ at low doping concentrations and they differentiate afterwards (Table S3, Figure 4a). The spin-only $\mu_{eff.cal}$ are underestimations of the observed moments, suggesting that the orbital angular momentum exerts some influence for the Mn cases. In Figure 4b, for the Mn-doped double perovskite $Sr_2Ga_{1-x}Mn_xSbO_6$ series, the observed magnetization $M$ increases with the doping level $x$, but the increments gradually reduce; hence the trend line shows the sign of leveling off at higher doping concentrations. In contrast to the Mn-doped double perovskite, $M$ for the disordered single perovskite $LaGa_{1-x}Mn_xO_3$ series exhibits a clear, positive linear relationship with respect to the doping level $x$. The observed magnetizations $M$ of the Mn-doped single perovskite and double perovskite are quite close at low doping concentrations, but clearly diverge afterwards, with the difference increasing at higher doping levels. For the same Mn-doping concentration, the $T_c$ of the ordered double perovskite (end member $T_c \approx 170(6)$ K) is significantly lower than that of the disordered single perovskite (end member $T_c \approx 200(6)$ K), and the magnetic transitions in the single perovskite series are much broader than those of the double perovskite series.

A possible explanation of our overall magnetic observations may be that in ordered double perovskites, the Mn-O-M-O-Mn (M is a non-magnetic cation) super-exchange pathway dominates the magnetic behavior, while in the disordered single perovskite, in addition to some Mn-O-M-O-Mn magnetic-exchange pathways, Mn-O-Mn pathways are also feasible, which cannot happen in the double perovskite except at defects. This pathway clearly should have a stronger magnetic coupling strength, as non-magnetic cations (i.e., the $Sb^{5+}$) are not involved. Mn-O-Mn super-exchange pathways are more likely to appear at higher doping levels in the single perovskite and thus Mn-O-Mn coupling in the disordered single perovskite series contributes more to the magnetic behavior. This accounts for the greater differences between the ordered and disordered perovskite series in both $T_c$ and the saturated ferromagnetic moment at the Mn-rich end of each series. This kind of suppression of the magnetic state has been widely observed in perovskites and their derivatives.[46–49]

*2.2.2 Calculated Band Structures and Characterization of Band Gaps*

The calculated band structures of the $Sr_2GaSbO_6$ and $Sr_2MnSbO_6$ double perovskites are shown in **Figure 5**. For $Sr_2GaSbO_6$ a calculated direct bandgap of ~ 1 eV is clearly observed at the Γ point while an indirect one (~3 eV) is seen between the Γ and N points; a nearly flat band exists between the Γ and X points (Figure 5a). The calculated band structure for ferromagnetic $Sr_2MnSbO_6$, shown in Figure 5b, is different. (The spin structure for the calculations was set to that found in the literature[37,38].) Firstly, it shows a bandgap for both spin directions. An indirect small bandgap (~0.1 eV) emerges between the Γ and P points for up spin while a much larger indirect one (~2 eV) shows up between Γ and X point for down spin. Thus for doping by either holes or electrons this material can be expected to be a spin polarized ferromagnetic semiconductor below its magnetic ordering temperature. The evolution of the bandgaps of the $Sr_2Ga_{1-x}Mn_xSbO_6$ series that we observe in the diffuse reflectance spectrum (**Figure 6**a) is consistent with our calculations of the electronic band structures.

The pseudo-absorbance, transferred from reflectance using the Kubelka-Munk function, is plotted against photon energy (eV) in Figure 6a for the doped double perovskite series. The optical band gaps were then analyzed based on Tauc plots (Table S4). For the undoped parent phase $Sr_2GaSbO_6$, the band gap is found to be 3.52 eV if the indirect-transition equation was used, while it is 3.83 eV if the direct-transition equation was applied instead. The band structure calculations (from both the literature[36] and our own) suggest this phase should have a direct band gap around 1.0 eV (Figure 5a), consistent with the fact that DFT calculations of band gaps for main group element compounds like this one are generally too low.

The size of band gap decreases with increasing doping level $x$ in the $Sr_2Ga_{1-x}Mn_xSbO_6$ system, accompanied by the samples' darker appearances. The band gaps are significantly suppressed upon doping, even 10% Mn-doping reduces the band gap to 1.15 eV (Figure 6c). The detailed absorption behavior is complicated in the Mn-doped double perovskite series, as double absorption features appear in the spectrum for $x = 0.1$ and 0.2, but for $x = 0.3$ and 0.4, the absorption in the higher energy regime (near 3 to 4 eV) has been significantly suppressed compared to earlier compositions. For $0.5 \leq x \leq 0.9$, these two absorptions seem to overlap and merge into a single absorption (lower than 1 eV). This can be attributed to the different band structures of the two end members, $Sr_2GaSbO_6$ and $Sr_2MnSbO_6$ (Figure 5). For low Mn concentrations, the materials exhibit absorption features from both $Sr_2GaSbO_6$ and $Sr_2MnSbO_6$, resulting in a double absorption behavior, then, with elevated Mn concentrations, the $Sr_2MnSbO_6$ featured absorption becomes dominant and thus leads to a major indirect transition lower than 1 eV. The optical gap for this ferromagnetic semiconductor, which should display polarized electronic states near the Fermi energy, is clearly continuously variable in the semiconductor regime under 2 eV. The band gaps measured by diffuse reflectance spectroscopy serve as strong evidence that the Mn-doped $Sr_2Ga_{1-x}Mn_xSbO_6$ double perovskites are semiconductors with tunable band gaps.

For the purpose of summarizing the systematic changes in ferromagnetic ordering temperatures $T_c$ of this series of materials, the magnetic susceptibility ($M/H$), modified by a small $\chi_0$ is plotted against temperature $T$, hence their first derivatives were taken to yield the embedded plot (Figure 6b). The temperature at which the $d(\chi_T - \chi_0)/dT$ curve deviates from the horizontal line around $y = 0$ is considered to be the $T_c$, as below this temperature, the Curie-Weiss law is no longer obeyed (also corresponds to the temperature at which the $1/\chi$ curve deviates from linearity in the Supporting Information (SI) Figures S3 – S16). Hence, $Sr_2Ga_{1-x}Mn_xSbO_6$ is found to be a ferromagnetic semiconductor, with ferromagnetic ordering temperature $T_c$ and size of band gap easily tuned by the Mn-doping concentration (Figure 6c).

## 3. Conclusions

Previous electronic structure calculations suggested to us that the B-site-cation-ordered magnetically silent oxide double perovskite $Sr_2GaSbO_6$ should be a semiconductor with a direct band gap of around 1 eV.[36] To make this material exhibit ferromagnetism, hence becoming a potential candidate material for spintronic devices, $Mn^{3+}$ was selected to partially substitute for $Ga^{3+}$. The doped double perovskite $Sr_2Ga_{1-x}Mn_xSbO_6$ materials crystallize in the same space group ($I4/m$) as their undoped parent phase $Sr_2GaSbO_6$, with the $Ga^{3+}/Mn^{3+}$ cations completely ordered with $Sb^{5+}$. Magnetization data collected from the doped phases show that the Mn-doped materials exhibit the desired ferromagnetic interactions, which become stronger with increasing Mn content. The ferromagnetic ordering temperature $T_c$ increases towards the Mn-rich end of the solid solution. In addition to the magnetic properties, the Mn-doped materials all have significantly smaller band gaps compared to their undoped parent phase $Sr_2GaSbO_6$, with the size of band gaps decreasing with elevated doping level $x$. The calculations suggest that these materials, in addition to being ferromagnetic and semiconducting, should also display highly spin polarized electronic states near $E_F$; an important characteristic for use in spin selective next-generation electronic devices.[50] Our results show that tunable ferromagnetic semiconducting double perovskites are viable materials for implementation in next-generation electronic devices.

**Experimental**

Approximately 0.5 g polycrystalline powder samples of the parent phase $A_2BB'O_6$ ordered double perovskite $Sr_2GaSbO_6$ and its doped phases $Sr_2Ga_{1-x}Mn_xSbO_6$ ($0.1 \leq x \leq 0.9$) were synthesized by a conventional solid state reaction method. Polycrystalline powder samples of the well-known B-site disordered single perovskite $LaGa_{1-x}Mn_xO_3$ ($0.1 \leq x \leq 0.5$) were also synthesized by this approach. Stoichiometric metal ratios of $SrCO_3$ (Alfa Aesar, 99.99%), $La_2O_3$ (Alfa Aesar, 99.99%, dried at 900 °C), $Ga_2O_3$ (Alfa Aesar, 99.999%), $Sb_2O_5$ (Alfa Aesar, 99.998%) and $Mn_2O_3$ (Alfa Aesar, 99.99%) were ground together using an agate mortar and pestle, and then transferred into an alumina crucible. These reaction mixtures were first slowly (1 °C/min) heated to 1000 °C in air and held overnight to decompose the carbonate, and then were directly annealed in air at 1300 – 1500 °C (3 °C/min) for 3 periods of 72 hours with intermittent grindings. The reaction progress was monitored using laboratory X-ray powder diffraction data collected at room temperature on a Bruker D8 FOCUS diffractometer (Cu Kα) over a 2θ range between 5° and 70°. Once the reactions were complete, laboratory XRD data with much better statistical significance, covering a 2θ range between 5° and 110°, were collected from each sample. Lattice parameters, atomic positions, and atomic displacement factors using these data were determined by the Rietveld method[51] using the GSAS-II program.

Approximately 3 g polycrystalline powder samples of the parent phase $Sr_2GaSbO_6$ and the 10%-doped phase $Sr_2Ga_{0.9}Mn_{0.1}SbO_6$, used for the neutron diffraction experiments, were also synthesized via the above approach. Time-of-flight (TOF) neutron powder diffraction data were collected at Oak Ridge National Laboratory's Spallation Neutron Source, POWGEN beamline, using neutron beam with a center wavelength of 0.8 Å at 300 K. Structures from this data were determined by the Rietveld method[51] using the GSAS-II program.

The magnetization data were collected using the vibrating sample magnetometer (VSM) option of a Quantum Design Physical Property Measurement System (PPMS). Temperature-dependent magnetization ($M$) data were collected from finely ground powders in an applied field ($H$) of 1000 Oe. The magnetic susceptibility $\chi$ was defined as $M$ (in emu)/$H$ (in Oe). Field-dependent magnetization data between $H$ = 90000 Oe and –90000 Oe were collected at $T$ = 300 K and 2 K.

The diffuse reflectance spectra were collected from powder samples at ambient temperature on a Cary 5000i UV-VIS-NIR spectrometer equipped with an internal DRA-2500 integrating sphere. The data were converted from reflectance to pseudo absorbance using the Kubelka–Munk method, and values of band transitions were calculated from Tauc plots.[52] The particle morphology of selected compositions was investigated using an FEI XL30 field-emission gun scanning electron microscope (SEM) equipped with an Oxford X-Max 20 energy-dispersive X-ray spectroscope (EDX) running on AZtec software.

The band structures and electronic densities of states (DOS) were calculated using the WIEN2k program. The full-potential linearized augmented plane wave (FP-LAPW) method with local orbitals was used.[53,54] Electron correlation was treated via the generalized gradient approximation.[55] The conjugate gradient algorithm was applied, and the energy cutoff was set at 500 eV. Reciprocal space integrations were completed over a 6×6×4 Monkhorst-Pack $k$-point mesh.[56] Spin-orbit coupling (SOC) effects were only applied for the Sb atom. Spin-polarization (ferromagnetism with the moment oriented in the (001) direction) was only employed for the Mn atoms. The lattice parameters of $Sr_2GaSbO_6$ were obtained from M. Lufaso *et al.*[39] For $Sr_2MnSbO_6$, the crystallographic data were retrieved from the Materials Project database.[57] The calculated total energy converged to less than 0.1 meV per atom.

**Supporting Information**

The Supporting Information is available from the Wiley Online Library or from the author.

**Acknowledgements**

This research was primarily done at Princeton University, supported by the US Department of Energy, Division of Basic Energy Sciences, grant number DE-FG02-98ER45706. A portion of this research used resources at the Spallation Neutron Source, a DOE Office of Science User Facility operated by the Oak Ridge National Laboratory. The authors acknowledge the use of Princeton's Imaging and Analysis Center, which is partially supported through the Princeton Center for Complex Materials (PCCM), a National Science Foundation (NSF)-MRSEC program (DMR-2011750).

**Conflict of Interest**

The authors declare no conflict of interest.

**Table 1.** Structural parameters and crystallographic positions from the refinement of neutron powder diffraction data collected from $Sr_2Ga_{0.9}Mn_{0.1}SbO_6$ at 300 K.

| Atoms | x/a | y/b | z/c | S.O.F. | $U_{iso}$ equiv. (Å$^2$) |
|---|---|---|---|---|---|
| Sr1 | 0 | 0.5 | 0.25 | 1 | 0.00656 |
| Ga1 | 0 | 0 | 0 | 0.9032(1) | 0.00349 |
| Sb1 | 0.5 | 0.5 | 0 | 1 | 0.00126 |
| O1 | 0 | 0 | 0.2495(6) | 1 | 0.01000 |
| O2 | 0.2224(6) | 0.2777(2) | 0 | 1 | 0.00775 |
| Mn1 | 0 | 0 | 0 | 0.0968(1) | 0.00349 |

$Sr_2Ga_{0.9}Mn_{0.1}SbO_6$ space group $I4/m$ (#87)
Formula weight: 461.24 g mol$^{-1}$, Z = 2
$a$ = 5.54094(4) Å, $c$ = 7.93548(4) Å, Volume = 243.635(3) Å$^3$
Radiation source: time of flight neutrons
Temperature: 300 K
$wR$ = 4.179%; $GOF$ = 3.54

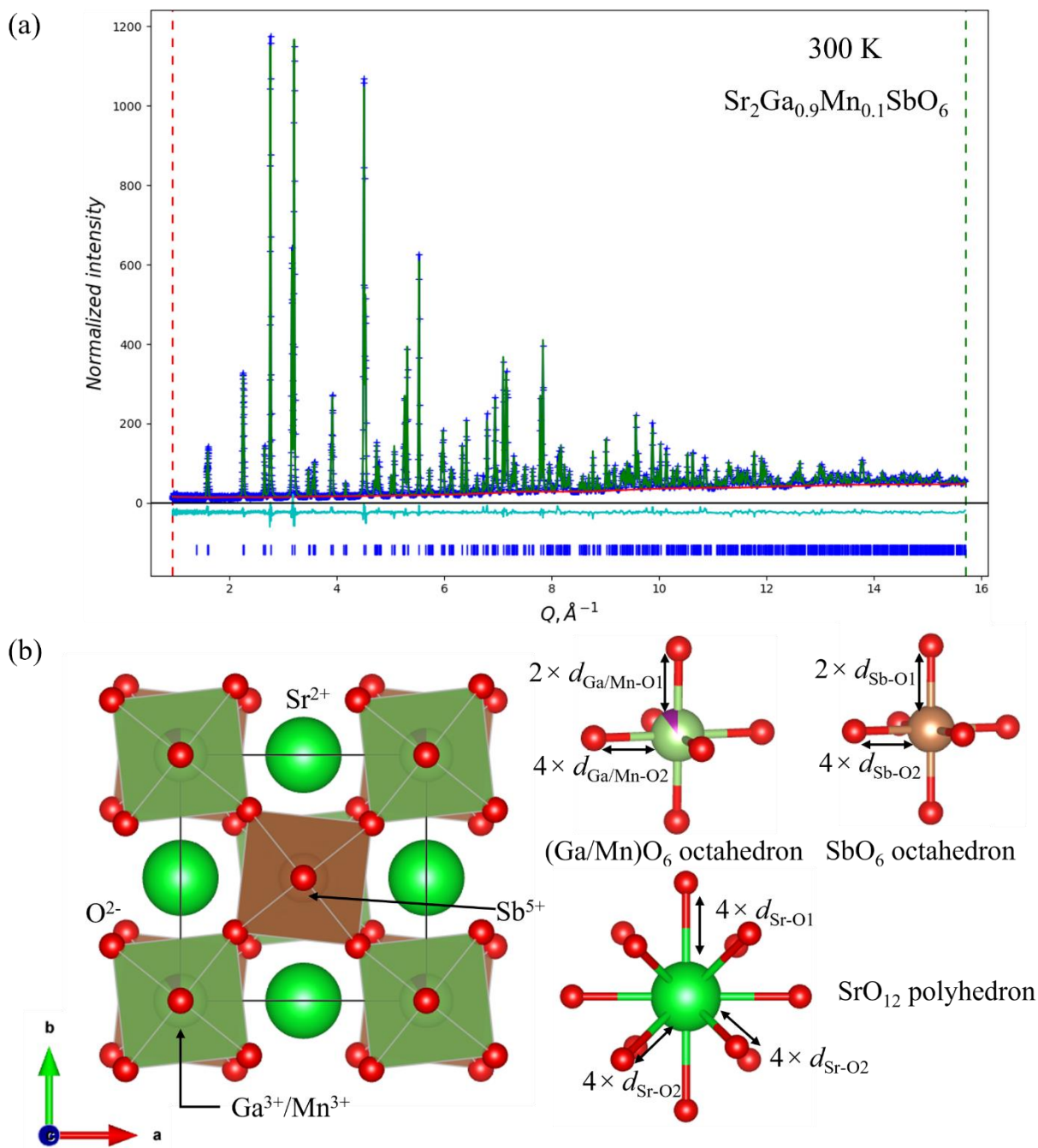

**Figure 1.** (a) Observed (blue), calculated (green) and difference (cyan) plots from the Rietveld refinement of $Sr_2Ga_{0.9}Mn_{0.1}SbO_6$ (space group $I4/m$) against neutron powder diffraction data at 300 K; (b) The structural model of $Sr_2Ga_{0.9}Mn_{0.1}SbO_6$, and the selected bonding environment of the $Ga_{0.9}Mn_{0.1}O_6$ octahedron, $SbO_6$ octahedron and the $SrO_{12}$ polyhedron. The noted bond lengths are listed in Table S2.

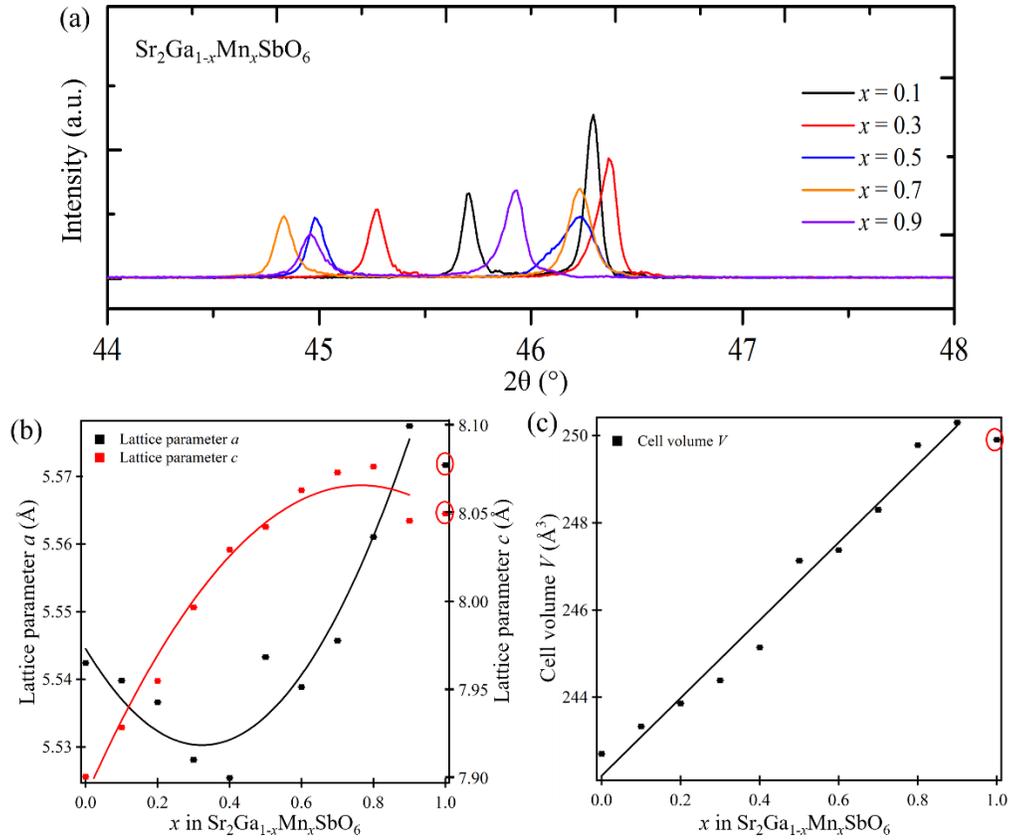

**Figure 2.** (a) Stacked lab X-ray diffraction patterns for selected compositions of $Sr_2Ga_{1-x}Mn_xSbO_6$ series; (b) lattice parameter *a* and *c*, (c) and cell volume *V* (error bars are smaller than the data points) plotted for each composition of the $Sr_2Ga_{1-x}Mn_xSbO_6$ series. (Data points in red circles are extracted from the all-Mn end member $Sr_2MnSbO_6$ in literature, which does not fall in line with the materials reported here.)

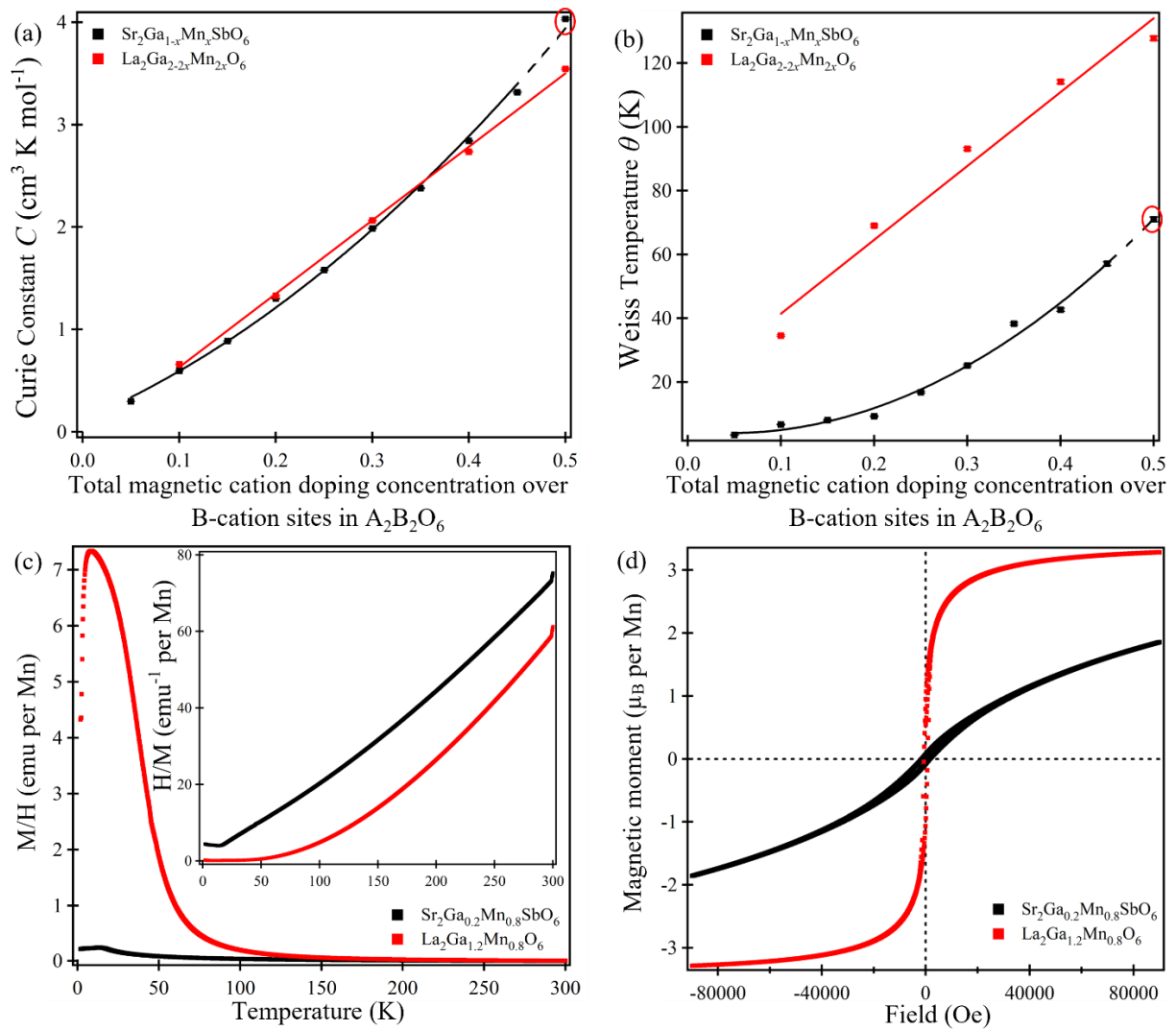

**Figure 3.** (a) The Curie constant and (b) the Weiss temperature (error bars are smaller than the data points) extracted from the fitting of paramagnetic susceptibility to the Curie-Weiss law for each composition of the $Sr_2Ga_{1-x}Mn_xSbO_6$ series and the $LaGa_{1-x}Mn_xO_3$ series (scaled based on its equivalent double perovskite formula); (c) the magnetic susceptibility $\chi$ ($M/H$) against temperature T, with $1/\chi$ against T plots inset; and (d) the magnetic moment $M$ against field $H$ for double- and single-perovskite with 40% total Mn-concentration over B-cation sites in $A_2B_2O_6$ (late members in each series). (Data points in red circles are extracted from the all-Mn end member $Sr_2MnSbO_6$ in literature.)

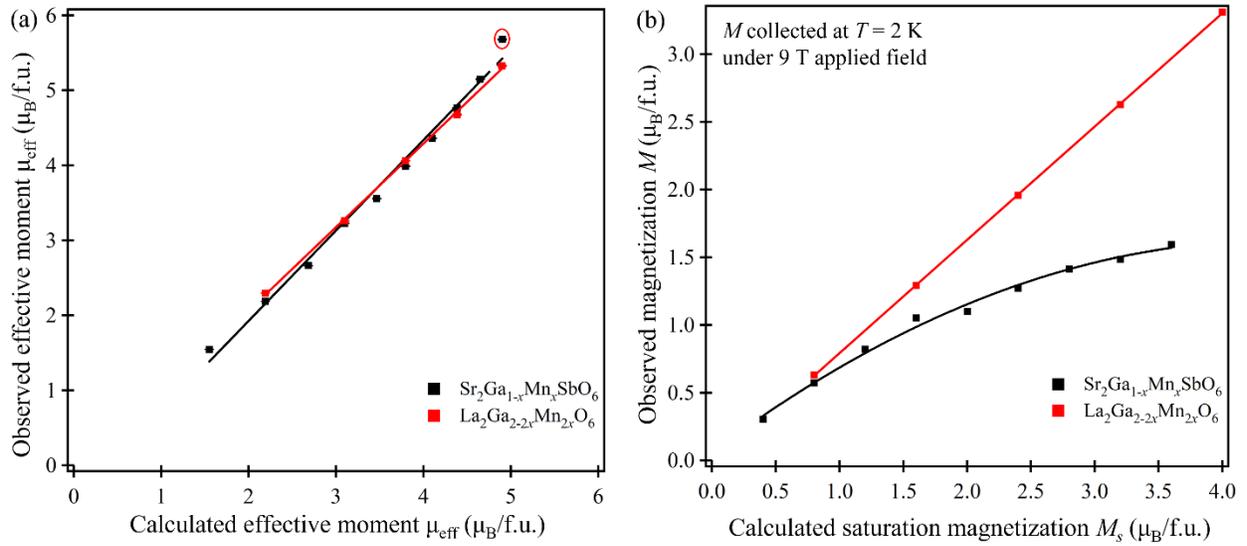

**Figure 4.** (a) The observed effective moment per formula unit plotted against the effective moment per formula unit calculated from the spin-only contribution; (b) the observed magnetization at 2 K under an applied field of 9 T (error bars are smaller than the data points) plotted against the calculated saturation magnetization for each composition of the $Sr_2Ga_{1-x}Mn_xSbO_6$ series and the $LaGa_{1-x}Mn_xO_3$ series (scaled based on its equivalent double perovskite formula). (Data point in the red circle is extracted from the all-Mn end member $Sr_2MnSbO_6$ in literature.)

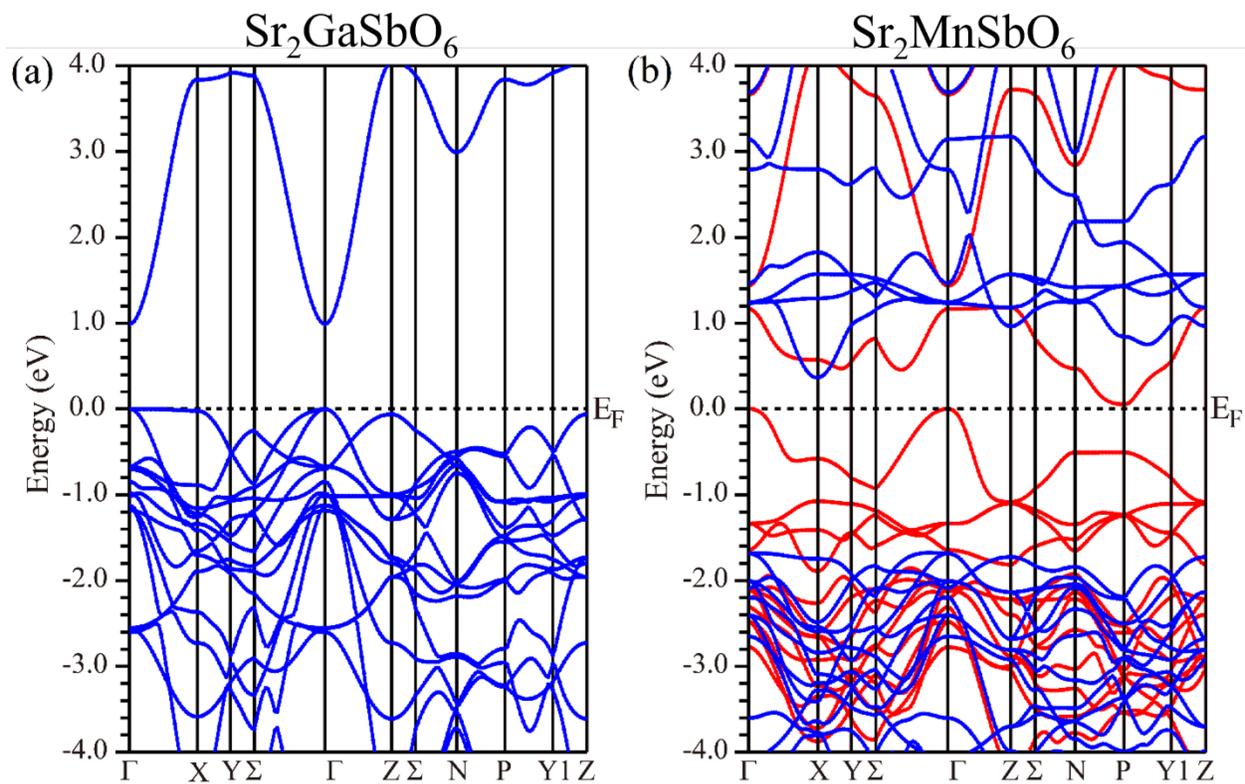

**Figure 5.** Calculated band structures and electronic density of states (DOS) of (a) $Sr_2GaSbO_6$ and (b) $Sr_2MnSbO_6$ (red for up spins and blue for down spins).

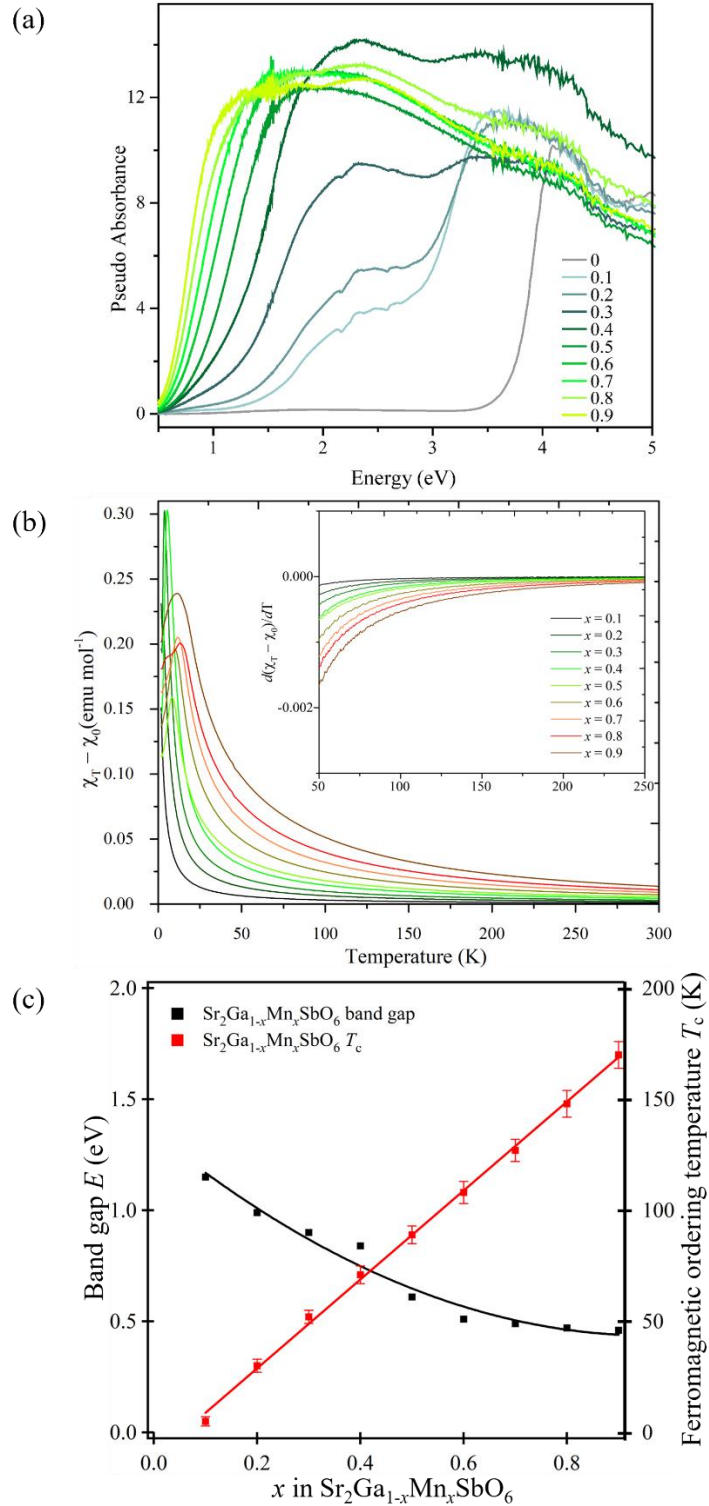

**Figure 6.** (a) The diffuse reflectance spectra; (b) ($\chi_T - \chi_0$) plotted against temperature, with the first derivative plot embedded and (c) the band gaps from Tauc plots obtained by using an indirect transition equation together with the ferromagnetic ordering temperature $T_c$ plotted against the doping level $x$ in the $Sr_2Ga_{1-x}Mn_xSbO_6$ series.